\documentclass[a4paper,fleqn,usenatbib]{mnras}

\usepackage{txfonts}

\usepackage{booktabs}

\usepackage[T1]{fontenc}
\usepackage{ae,aecompl}

\usepackage{graphicx}	
\usepackage{amssymb}	

\title[Morpho-kinematical structure of NGC\,1514]{The morpho-kinematical structure and chemical abundances of the complex planetary nebula NGC\,1514}

\author[A. Aller et al.]{
A. Aller$^{1, 2}$\thanks{E-mail: alba.aller@cab.inta-csic.es},
R. V\'azquez$^{3}$,
L. Olgu\'in$^{4}$,
L.~F. Miranda$^{5}$
and M. Ressler$^{6}$
\\ 
$^{1}$Departamento de Astrof\'{\i}sica, Centro de Astrobiolog\'{\i}a (INTA-CSIC), PO Box 78, Villanueva de la Ca\~nada (Madrid) E-28691, Spain\\
$^{2}$Spanish Virtual Observatory, Spain\\
$^{3}$Instituto de Astronom\'{\i}a, Universidad Nacional Aut\'onoma de
M\'exico, Apdo. Postal 877, 22800 Ensenada, B.C., Mexico\\
$^{4}$Departamento de Investigaci\'on en F\'{\i}sica, Universidad de Sonora,
Blvd. Rosales Esq. L.D. Colosio, Edif. 3H, 83190 Hermosillo,\\ Son. Mexico\\
$^{5}$Instituto de Astrof\'{\i}sica de Andaluc\'{\i}a - CSIC, C/ Glorieta de la Astronom\'{i}a s/n, E-18008 Granada, Spain\\
$^{6}$Jet Propulsion Laboratory, California Institute of Technology, 4800 Oak Grove Drive, Pasadena, CA 91109, USA\\
}

\date{Accepted XXX. Received YYY; in original form ZZZ}

\pubyear{2015}

\begin{document}
\label{firstpage}
\pagerange{\pageref{firstpage}--\pageref{lastpage}}
\maketitle

\begin{abstract}
We present high-resolution, long-slit optical spectra and images of the planetary nebula NGC\,1514. The 
position velocity maps of the [O\,{\sc iii}] emission line reveal complex kinematics with multiple structures. A morpho-kinematical analysis suggests an inner shell, originally spherical and now distorted by several bubbles, and an attached outer shell. The two well-defined, mid-infrared rings of NGC\,1514 are not detected in our high-resolution, long-slit spectra, which prevented us from doing a kinematical analysis of them. Based exclusively on their morphology, we propose a barrel-like structure to explain the rings. Several ejection processes have been possibly involved in the formation of the nebula although a time sequence is difficult to establish with the current data. We also analyze intermediate-resolution, long-slit spectra with the goal of studying the physical parameters and chemical abundances of NGC\,1514. The nebular spectra reveal a moderate-excitation nebula with weak
emission lines of  [Ar\,{\sc iii}], [Ne\,{\sc iii}], He\,{\sc i} and He\,{\sc ii}. No [N\,{\sc ii}] neither other low-excitation emission lines are detected. We found an electron temperature around 14000\,K in the gas and an electron density in the range of 2000 and 4000 cm$^{-3}$.

\end{abstract}




\section{Introduction}

Among the planetary nebula (PN) zoo one can find a great diversity of morphologies. In general terms, 
PNe were originally classified as round, elliptical, bipolar, and irregular. However, this is a very broad scheme and not all the PNe can be clearly classified in one of these groups. Observations at higher angular resolution of PNe have revealed more complex features and microstructures,
which led to other morphological types like quadrupolar, multipolar and point-symmetric \citep[see 
e.g.,][]{Manchado1996a, Manchado1996b}. In addition, other structures like jets, knots, and rings are also
identified in many PNe \citep[see e.g.][]{Miszalski2009b, Boffin2012, Guerrero-Miranda2012}, which complicates enormously the classification and, therefore, the study of the physical processes that may be involved in their shaping. 
In addition, some PNe present different morphologies or structures when observed at different wavelengths \citep[see e.g.,][]{Ramos-Larios2008}.

NGC\,1514 ($\alpha$ $\! = \!$ 04$^ {\rm h}$\,09$^ {\rm m}$\,16$\fs$9,
$\delta$ $\! = \!$ +30$^{\circ}$\,46$'$\,33$''$, equinox 2000.0) is a PN with a complex morphology that does not clearly fit in the ``defined" morphological types. In the optical wavelength range, this nebula presents an irregular inner shell composed of several bubbles and a fainter attached outer shell. The nebula was classified as type II multiple-shell PN by \cite{Chu1987}. Furthermore, a pair of axisymmetric rings contained within the outer shell, are only visible in the mid-infrared wavelengths \citep{Ressler2010}. In addition to this complex morphology, this PN is also interesting for its {\it peculiar} central source (BD+30$^{\circ}$623). It belongs to the small group of PNe with an absorption spectrum of a cool central star \citep[spectral type A through K;][]{Lutz1977,Mendez1978}. These {\it peculiar} central stars should have a faint hot companion, which would be the responsible for the excitation of the PN. It is the case of BD+30$^{\circ}$623, with a typical A-type spectrum that shows signatures of a hot companion. The binarity of the central star has been further discussed in the literature \citep[see][for a review of the issue]{Aller2015b, Mendez2016} and BD+30$^{\circ}$623 was recently confirmed to be the widest binary central star to date, with an orbital period of $\sim$ 3300 days \citep{Jones2017}.

There is no doubt now that binary central stars play a key role in the formation of a remarkable number of PNe with non-spherical morphologies (see \citealt{Boffin-Jones2019} for an extensive review of the issue). In particular, the fraction of close binary central stars, i.e. those that have experienced the common envelope (CE) phase and have orbital periods between a few hours and several days, seem to represent at least 12 -- 21\% of the central stars of PNe, although this number may be a lower limit \citep{Miszalski2009a}. In addition, hydrodynamic simulations \citep[see e.g.,][]{Theuns1996, Edgar2008} have demonstrated that also wide binaries can influence the shape of PNe via mass transfer by stellar winds. Nevertheless, just a few PNe hosting wide binaries have been discovered so far, mainly due to the large observational efforts required.

Since it was first discovered by William Herschel in 1790, NGC\,1514 has been extensively studied to date.
However, the unique spatio-kinematical analysis of NGC\,1514 so far is that presented by \cite{Muthu-Anandarao2003}. 
They used an imaging Fabry-P\'erot spectrometer and concluded
 that the nebula consists of three structures: an inner ellipsoidal shell, a faint outer shell, and polar bright blobs embedded 
 inside the main shell that do not follow a bipolar morphology. However, the proposed model does not fully account for the structural complexity of NGC1514. In addition, \cite{Muthu-Anandarao2003} were unable to derive a systemic radial velocity for the nebula. High-resolution spectroscopy had not been 
  used so far and it is really needed to fully understand all the components and structures presented in the nebula. 

In this work, we present high-resolution, long-slit spectra and images of NGC\,1514 which allow us to describe 
in detail the internal kinematics and morphology of the nebula. In addition, we also present intermediate-resolution, long-slit spectra with the aim of deriving their physical conditions and chemical abundances.

\section[]{Observations}

\subsection{Optical imaging}

Figure\,1 shows the narrow-band [O\,{\sc iii}] image obtained on 2011 January 16 with the Calar Alto Faint 
Object Spectrograph (CAFOS) at the 2.2-m telescope on Calar Alto
Observatory\footnote{Data here reported were acquired at Centro Astron\'omico Hispano Alem\'an (CAHA) at Calar Alto operated jointly by Instituto de Astrof\'{i}sica de Andaluc\'{i}a (CSIC) and Max Planck Institut f\"ur Astronomie (MPG). Centro Astron\'omico Hispano en Andaluc\'ia is now operated by Instituto de Astrof\'isica de Andaluc\'ia and Junta de Andaluc\'ia.} (Almer\'{i}a,
Spain). A SITe 2k$\times$2k--CCD was used as detector, with a plate 
scale of 0.53\,arcsec pixel$^{-1}$ and a circular field of view of 16 arcmin in
diameter. The total exposure time was 1900\,s (1$\times$100\,s + 3$\times$600\,s) in [O\,{\sc iii}] filter
($\lambda_{\rm 0}$ = 5007 \AA, FWHM = 87 \AA). An H$\alpha$ image ($\lambda_{\rm 0}$ = 6563 \AA, FWHM = 15 \AA) was also obtained (1$\times$100\,s + 3$\times$600\,s)
during the same observing run. The H$\alpha$ image does not present significant differences with the [O\,{\sc iii}] one and is not shown here.
The seeing was $\simeq$1.5 arcsec. 
The images were reduced following standard procedures for direct images
within the {\sc iraf} and {\sc midas} packages.

\begin{figure}
\includegraphics[width=0.5\textwidth]{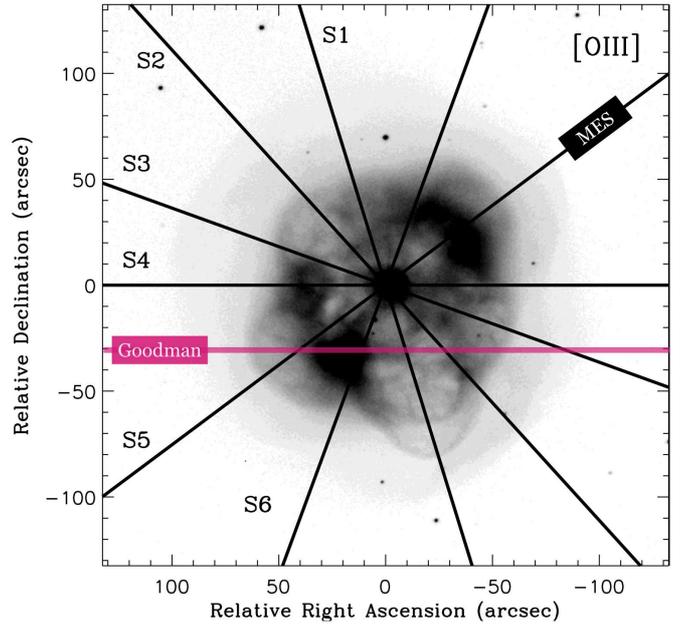}
  \vspace*{1pt}
  \caption{Gray-scale image of NGC\,1514 in the [O\,{\sc iii}] filter. Gray levels are
        linear. Slit positions used for high- and intermediate-, long-slit spectroscopy are overplotted in black and pink, respectively. 
(The width of the slits is not to scale).}
\label{fig:image_slits}
\end{figure}

As can be seen in Figure\,\ref{fig:image_slits}, NGC\,1514 presents a very complex morphology with a main shell composed of several bubbles and features. In addition, a diffuse attached shell, that will be referred hereafter as outer shell, is also recognized in the image. All these structures along with the pair of infrared rings discovered by \cite{Ressler2010} will be analyzed in detail in Sections \ref{sect:kinematics} and \ref{sect:shape}.

\subsection{High-resolution optical spectra}

High-resolution, long-slit spectra were obtained with the Manchester Echelle 
Spectrometer \citep[MES,][]{Meaburn2003} at the 2.1\,m telescope on San Pedro M\'artir Observatory
(OAN-SPM)\footnote{The  Observatorio Astron\'omico
  Nacional at the Sierra de San Pedro M\'artir (OAN-SPM) is operated by the
  Instituto de Astronom\'{\i}a of the Universidad Nacional Aut\'onoma de 
M\'exico.} during two observing runs: one on 2012 November 26-27 and the other on 2014 October 12-13.
A 2k$\times$2k Marconi CCD was used as detector in 4$\times$4 
binning mode (0.702\,arcsec pixel$^{-1}$). A $\Delta$$\lambda$ = 50 {\AA } filter was used to
isolate the [O\,{\sc iii}] emission line (114$^{\rm th}$
order), with a dispersion of 0.08\,$\AA$\,pixel$^{-1}$. The slit (6$'$ long, 2$''$ wide) was centered on 
the central star and spectra were obtained at six position angles (PAs): 17$^{\circ}$, 42$^{\circ}$, 70$^{\circ}$, 90$^{\circ}$, -53$^{\circ}$, and -20$^{\circ}$. These slits  (denoted S1, S2, S3, S4, S5, and S6, respectively)
are superimposed on the [O\,{\sc iii}] image in Fig.\,1. 
Exposure time was 1200\,s for the S1, S2 and S6 spectra and 1800\,s for S3, S4, and S5. A Th-Ar lamp was used
for wavelength calibration to an accuracy
of $\pm$1 km\,s$^{-1}$. The resulting spectral resolution (FWHM) is 12
km\,s$^{-1}$. Seeing was $\simeq$2$''$ during the observations. 

The spectra were cleaned of cosmic rays, de-biased, and wavelength calibrated using standard routines for long-slit spectroscopy
within the {\sc iraf} and {\sc midas} packages.

\subsection{Intermediate-resolution optical spectra}

Intermediate-resolution, long-slit spectra were obtained on 2017 December 28 with the Goodman High Throughput Spectrograph \citep{Clemens2004} mounted on the 4.1m Southern Astrophysical Research (SOAR) Telescope\footnote{Based on observations obtained at the Southern Astrophysical Research (SOAR) telescope, which is a joint project of the Minist\'{e}rio da Ci\^{e}ncia, Tecnologia, Inova\c{c}\~{o}es e Comunica\c{c}\~{o}es (MCTIC) do Brasil, the U.S. National Optical Astronomy Observatory (NOAO), the University of North Carolina at Chapel Hill (UNC), and Michigan State University (MSU).}, located on Cerro Pach\'on (Chile). We used the Red Camera equipped with a 4096$\times$4112 pixel, back-illuminated, e2v 231-84 CCD.
We used a 600 lines\,mm$^{-1}$ dispersion grating, giving a dispersion of 0.65\,$\AA$\,pixel$^{-1}$, and covering the 3500--6160 $\AA$ 
and 6300--8930 $\AA$ spectral ranges in the Blue and Red mode, respectively. The resulting spectral resolution is $\sim$4\,$\AA$. Two spectra were obtained with the Blue mode, with exposures times of 300\,s and 900\,s, and three in the Red one, with 300\,s each; all of them with the slit at PA 90$^{\circ}$ and at 30\,arcsec
southern of the central star. The slit width was 1\,arcsec and seeing was $\simeq$ 0.8\,arcsec during the observations.

The spectra were reduced following standard procedures for long-slit
spectroscopy within the {\sc iraf} packages. The
reduction included bias subtraction and flat-field
correction. Then, the spectra were wavelength calibrated, sky subtracted and,
finally, flux calibrated.

\begin{figure*}
\includegraphics[width=1.0\textwidth]{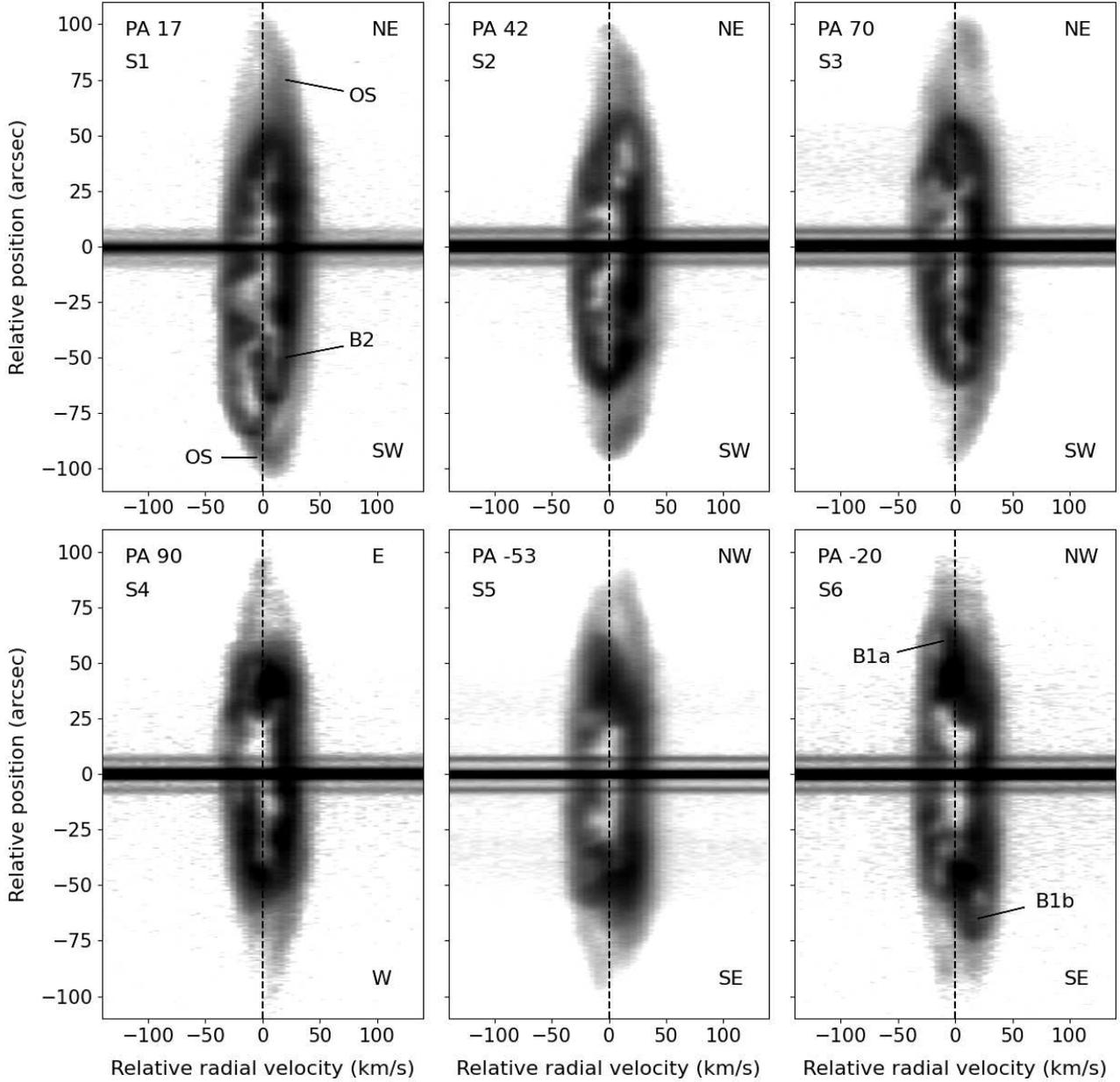}
  \vspace*{1pt}
  \caption{Logarithmic grey-scale, PV maps derived from the high-resolution, long-slit [O\,{\sc iii}] spectra of NGC\,1514
at six different PAs (upper left corner in each panel, see also Fig.\,1). The origin corresponds to the systemic velocity (see text) and position of the central star, as indicated by the stellar continuum. The two horizontal emission features
parallel to the continuum of the central star are a well-characterized reflection of the instrument. The main structures identified in the nebula are indicated in two of the panels (see also Table~\ref{table:structures} and Fig.~\ref{fig:sketch}).}
\label{fig:high_res_spectra}
\end{figure*}

\begin{figure}
\includegraphics[width=0.45\textwidth]{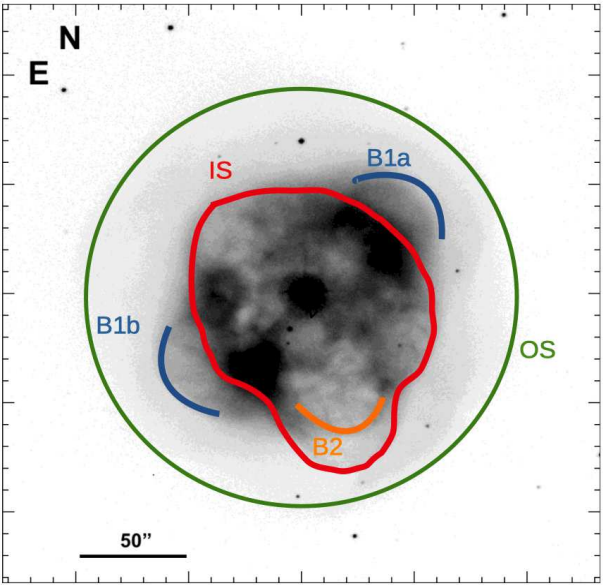}
  \vspace*{1pt}
  \caption{Sketch of the different structures of NGC\,1514 derived from the {\sc shape} model superimposed on the [O\,{\sc iii}] image.}
  \label{fig:sketch}
\end{figure}

\begin{figure*}
\includegraphics[width=0.32\textwidth]{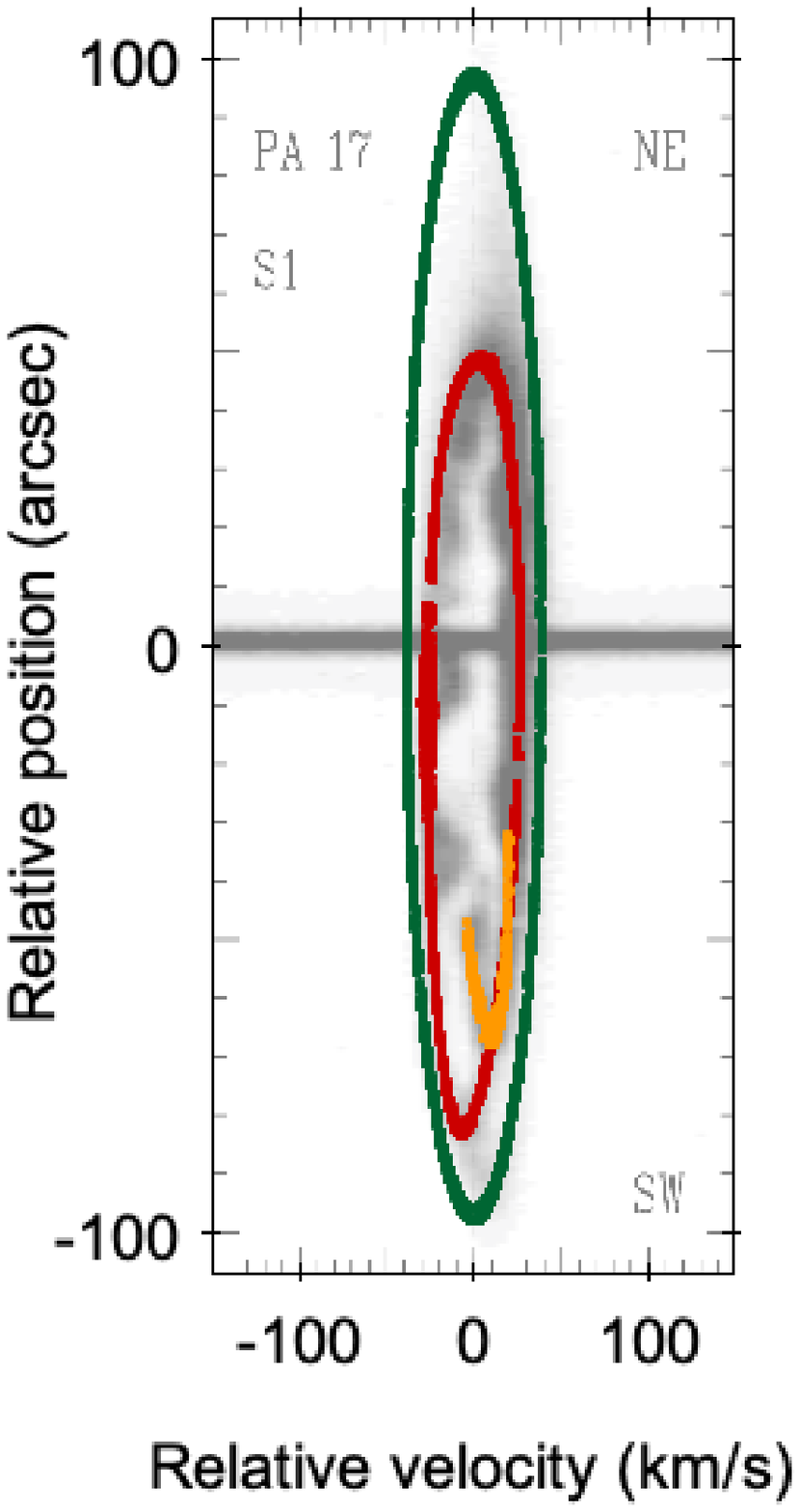}
\includegraphics[width=0.32\textwidth]{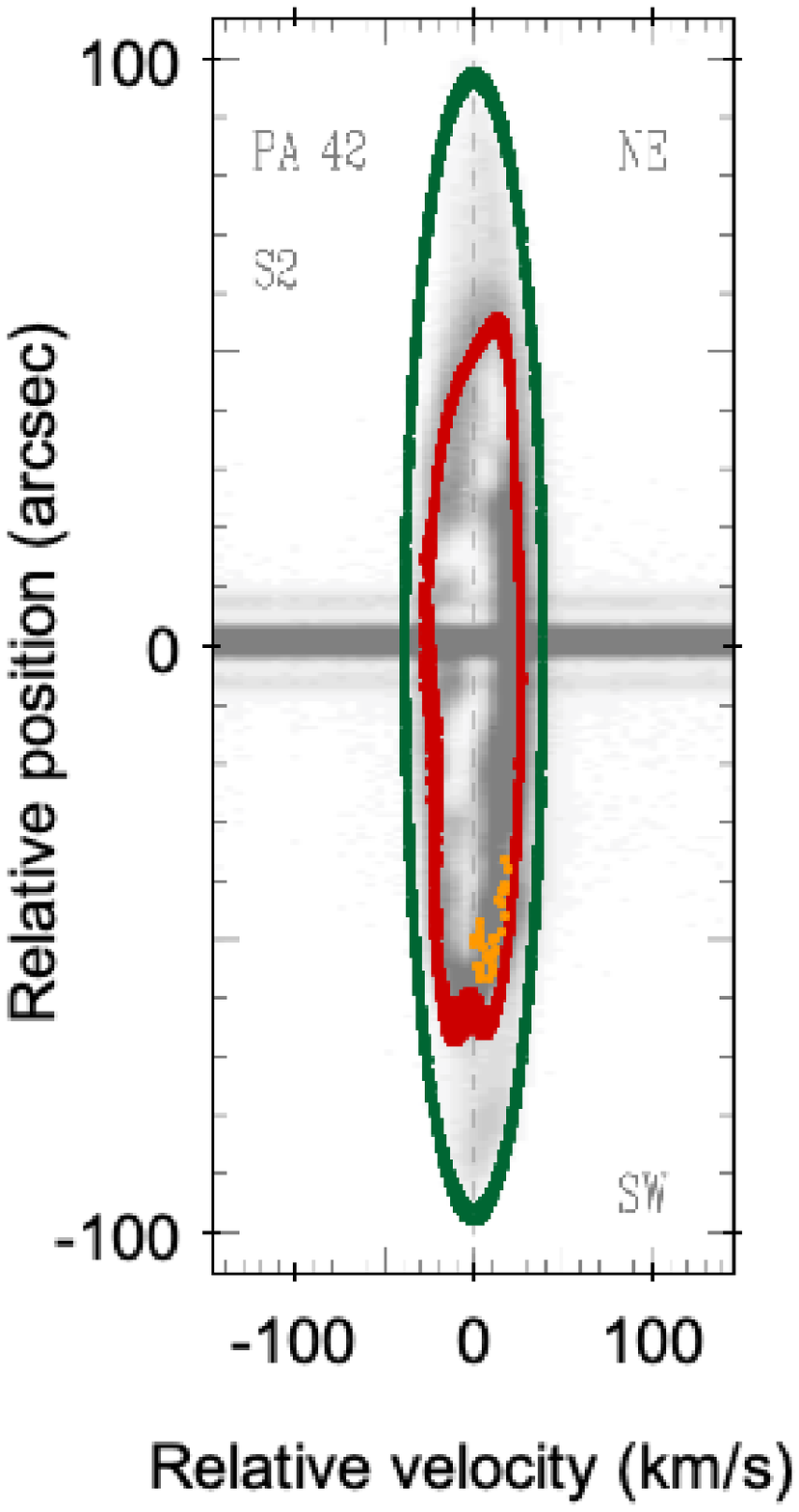}
\includegraphics[width=0.32\textwidth]{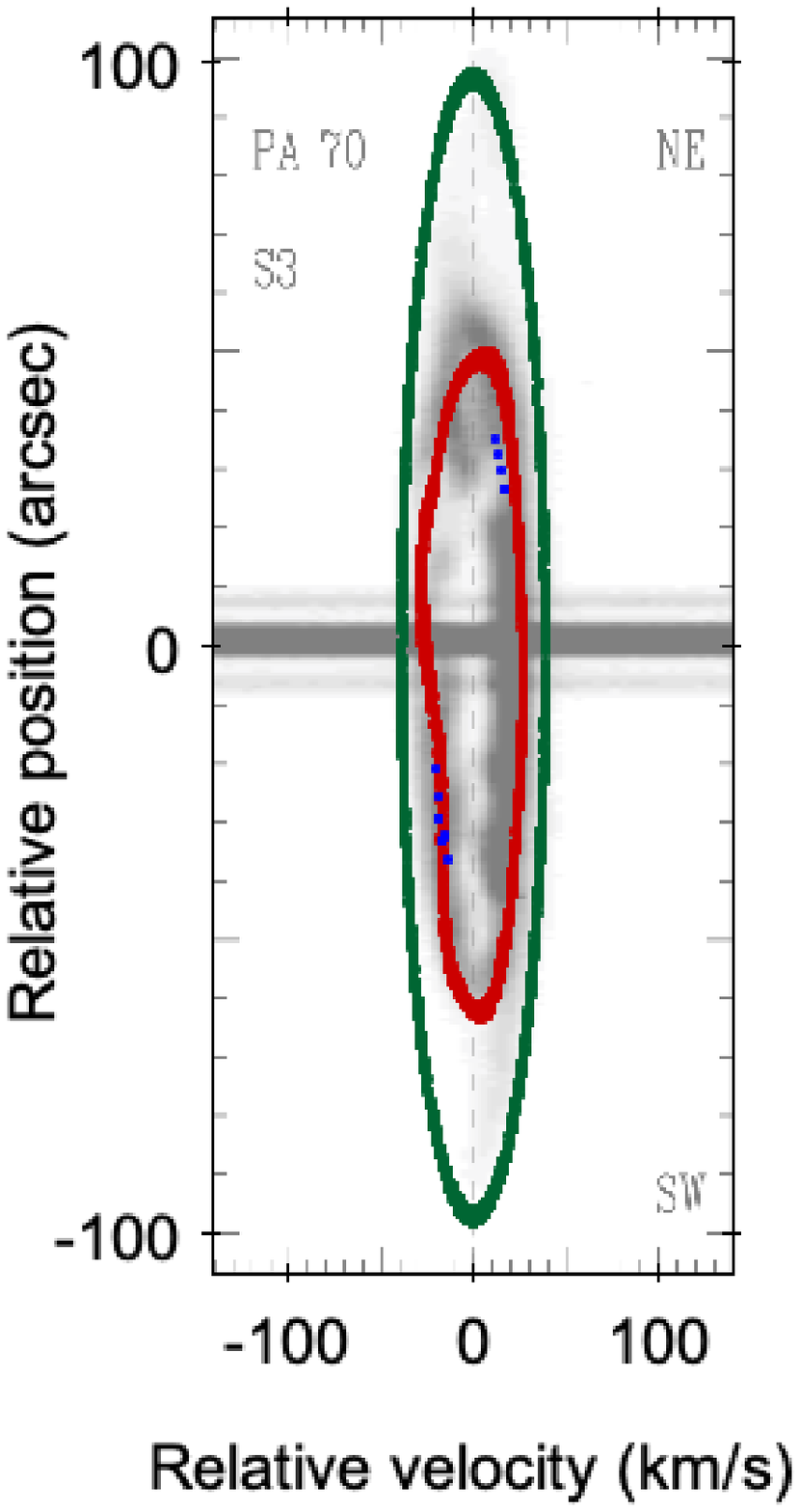}
\includegraphics[width=0.32\textwidth]{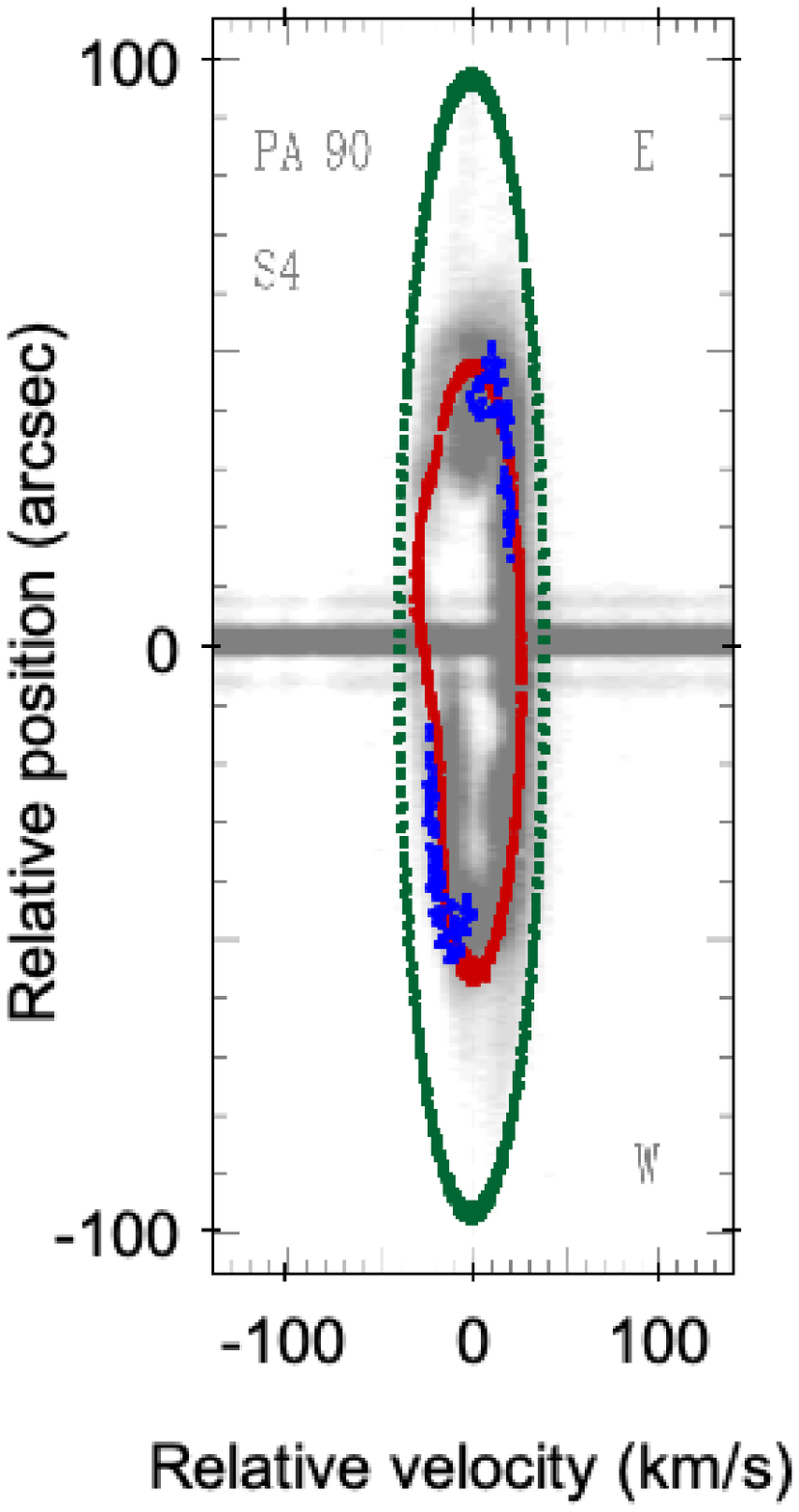}
\includegraphics[width=0.32\textwidth]{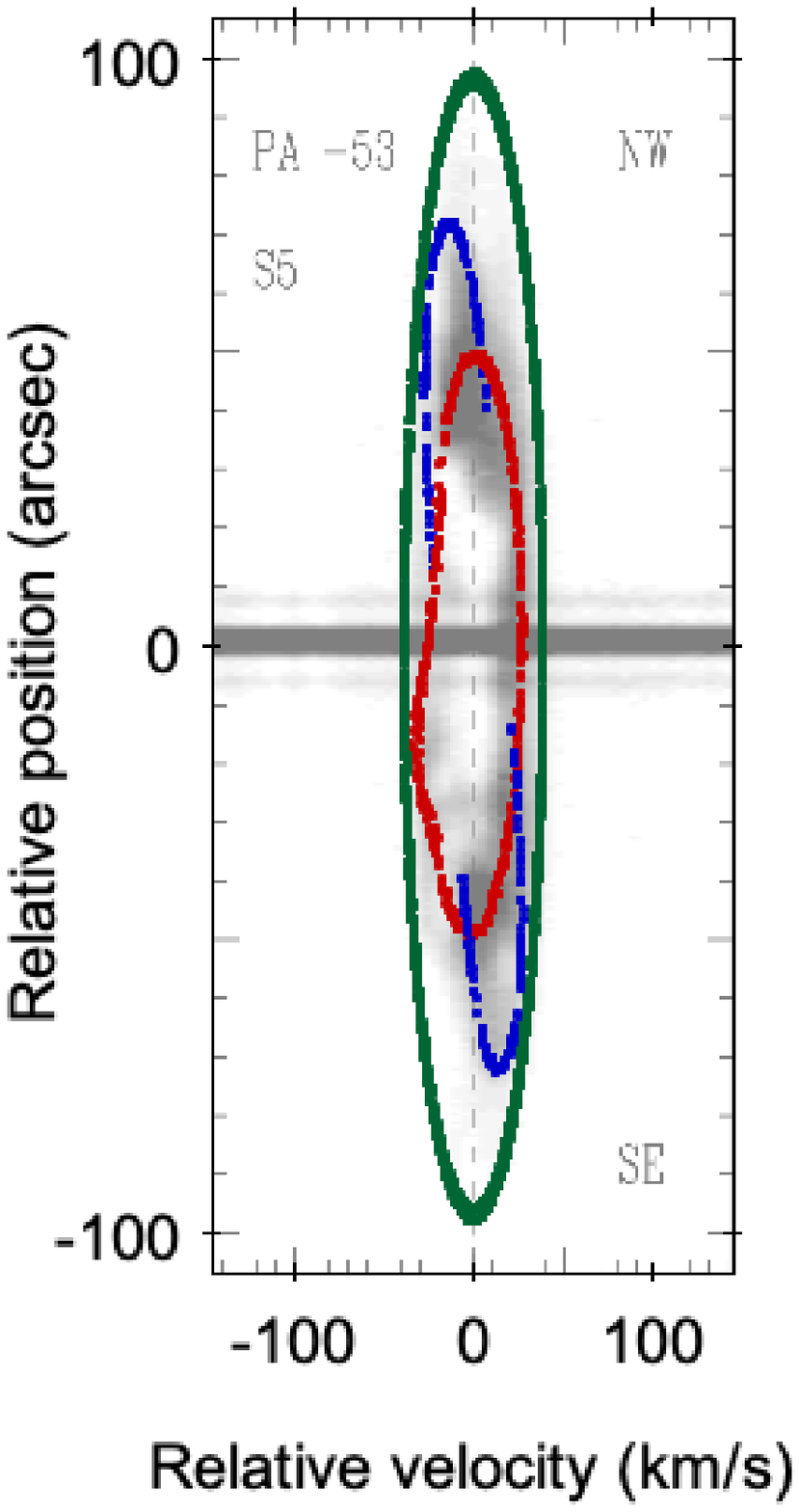}
\includegraphics[width=0.32\textwidth]{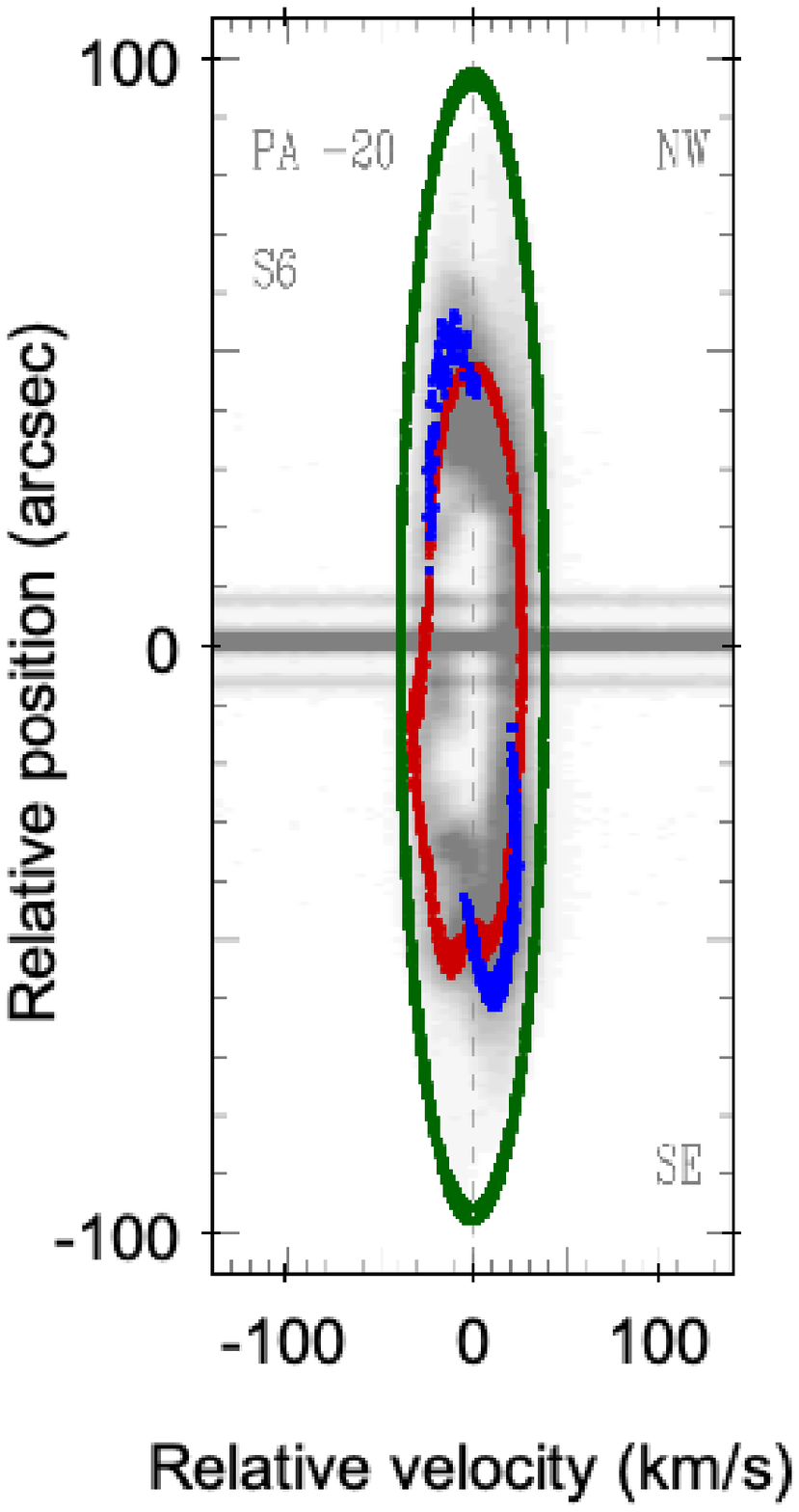}
  \vspace*{1pt}
  \caption{{\sc shape} synthetic PV-maps superimposed on the observed PV-maps of the [O\,{\sc iii}] line. Each color correspond to a different structure (see also Fig.~\ref{fig:sketch})}
  \label{fig:shape_spectra}
\end{figure*}

\section{Spatio-kinematical analysis}
\label{sect:kinematics}

Figure\,\ref{fig:high_res_spectra} shows position-velocity (PV) maps of the [O\,{\sc iii}] emission line at the six observed PAs. These PV maps allow us to reconstruct the true structure of the nebula and to impose some constraints on the shaping processes of NGC\,1514.

From the velocity centroid of the line emission features at all PAs we derive a heliocentric systemic velocity of $45\pm1\,{\rm km\,s^{-1}}$.
Previous measurements of the systemic velocity of NGC\,1514 range from 41.5$\pm$5.5\,km\,s$^{-1}$ to 91.0$\pm$9.0\,km\,s$^{-1}$ \citep[see][and references therein]{Schneider1983}. However, none of these measurements were derived from high resolution spectra, so an accurate value had not been determined so far. For the first time, we present a reliable value for the systemic velocity, which is also in a relatively good agreement with the systemic velocity of the binary central star system \citep[48.7$\pm$0.5\,km\,s$^{-1}$,][]{Jones2017}. 
Internal radial velocities will be quoted hereafter 
with respect to the systemic velocity derived in the present work. The origin of radial velocities in the PV 
maps is the systemic velocity and the origin for projected angular distances is the 
position of the central star, as given by the intensity peak of the stellar continuum
that is detected in all long-slit spectra.

The PV maps present very complex kinematics that are described in the following paragraphs. 

At a first look, the PV map at S1 shows a velocity ellipse with additional complex features inside. It is clearly asymmetrical in size with respect to the stellar position, extending up to $\sim$ 50$\arcsec$ towards the north-east (NE) and $\sim$ 82$\arcsec$ towards the south-west (SW), fitting very well the size of the nebula along this PA (see Fig.\,\ref{fig:image_slits}). The velocity ellipse is 
 slightly tilted with the northeastern part moving away from the observer and the southwestern part approaching to the observer. In addition, the image shows two bubbles in the SW part of the nebula, that are also recognized in the PV map, revealing that they are, indeed, two different structures with different kinematics. The smaller one, with the tip at $\sim$\,66\,\arcsec south of the stellar position, is redshifted, whereas the bigger one is blueshifted, with its vertex at $\sim82$\,\arcsec south the stellar position and coinciding with the end of the velocity ellipse. Inside the velocity ellipse, some bright features are also observed in the blueshifted part of the spectrum, much more noticeable towards the SW, with a wide distribution of velocities (from almost the systemic velocity up to the maximum radial velocity). These bright features coincide with ``clumps'' observed in the image (see Fig.\,\ref{fig:image_slits}). This could suggest that in these regions the inner shell is not completely empty but there is a concentration of material in these
 areas. Beyond the velocity ellipse, faint extended emission is also visible up to $\pm$100$\arcsec$ approximately, 
 being probably another velocity ellipse which corresponds to the faint outer shell observed in the image. This more diffuse velocity ellipse, which is visible in all the PV maps, appears slightly redshifted at the southwestern tip in the PV map at S1. However, the velocity ellipse seems to be centered at the systemic velocity in the rest of the PV maps.

The PV maps at S2 and S3 show similar features to each other. They both show a velocity 
ellipse of a size similar to the inner shell seen in Fig.\,\ref{fig:image_slits}. At S2, the velocity ellipse is tilted in the same way as the PV map at S1 (with the NE towards red and the SW towards blue), whereas the tips of the ellipse in S3 seem to be very close to the systemic velocity.  In the case of S3, an arc-like feature is also identified in the NE. Finally, as in the case of the PV map at S1, a second and fainter velocity ellipse is marginally detected beyond the main one in both PV maps, corresponding to the outer shell.

The appearance of PV map at S4 is slightly different from the previous ones. In this case, the velocity ellipse appears to have a flattened shape in the tips. In addition, the main characteristic is that a protrusion is clearly visible in the blueshifted part of the spectrum (specially prominent to the east of the stellar position). This ``bump'' would indicate the presence of a possible bubble in the direction of our line of sight that it is not recognized in the optical image. 

Finally, the PV maps at S5 and S6 present similar characteristics to each other, somewhat differing from the rest of the PAs. In these PV maps, the velocity ellipse appears disrupted, specially at the tips. The extremes of S5 clearly includes the line split corresponding to the bubbles seen in the image (that we will name structures B1a and B1b in Sect.\,\ref{sect:shape}). These bubbles are also identified in the PV map at S6, although they are not so clear since in this case the slit is only grazing the bubbles on their edges. 

All the PV maps reveal that the emission is considerably stronger in the redshifted part of the nebula than in the blueshifted one, which may suggest interaction with the interstellar medium.

\begin{table*}
	\centering
	\caption{Geometric and physical parameters of structures in {NGC\,1514}.}
	\label{tab:structures}
	\begin{tabular}{llccrcr} 
		\hline
		Name & Structure & Radius$^1$   & PA (\degr) & $i$ (\degr)  
		                                                       & $V_{\rm exp}$ & 
		                                                        Notes \\
		&   &  (arcsec)  &     & & ( km\,s$^{-1}$) & \\
		\hline
		B1a & Bubble 1a  & 78 & $-53$  & 63    & 42  & Aligned to S5; NW bubble.\\
		B1b & Bubble 1b  & 78 & +127   & 117  & 42  & Aligned to S5; SE bubble.\\
		B2   & Bubble 2    & 68 & +127   & 71    & 39  & Aligned to S1; SW small bubble.\\
		OS     & Outer Shell   & 98  & 0  & 0 & 39 & Spherical shell.\\
		IS       & Inner Shell$^2$   & 48 & 0 & 0 & 26 &  Irregular shell (originally spherical). \\
		\hline
		\label{parameters}
		\end{tabular}
		
		$^1$The number corresponds to the distance from the central star to the tip of the structure.\\ 
		$^2$The radius corresponds to that of the original sphere and $V_{\rm exp}$ to its expansion velocity. This sphere is distorted 
		by some `bumps'. 
\vskip 0.5 in
\label{table:structures}
\end{table*}

\begin{figure*}
\includegraphics[width=0.32\textwidth]{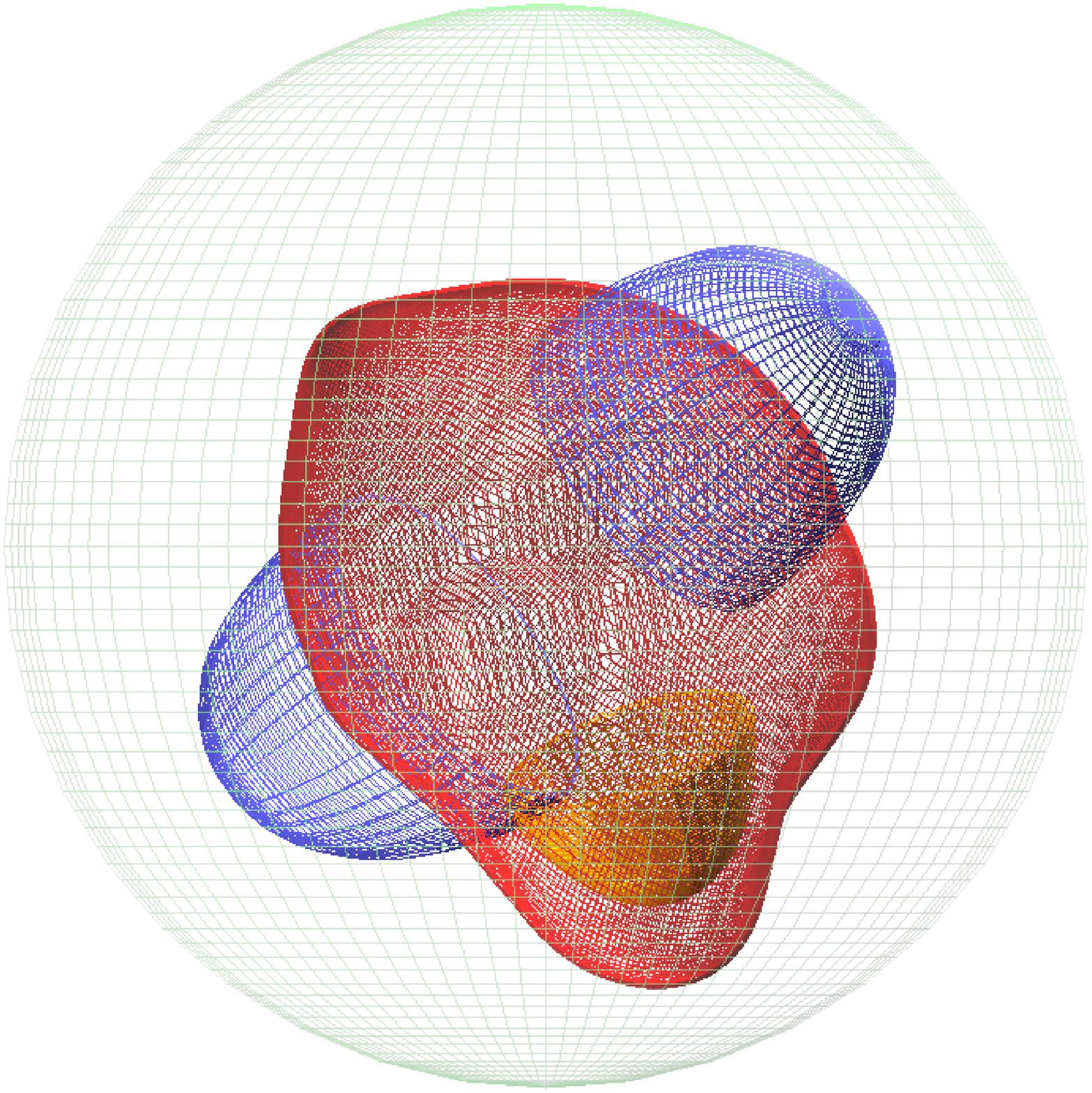}
\includegraphics[width=0.32\textwidth]{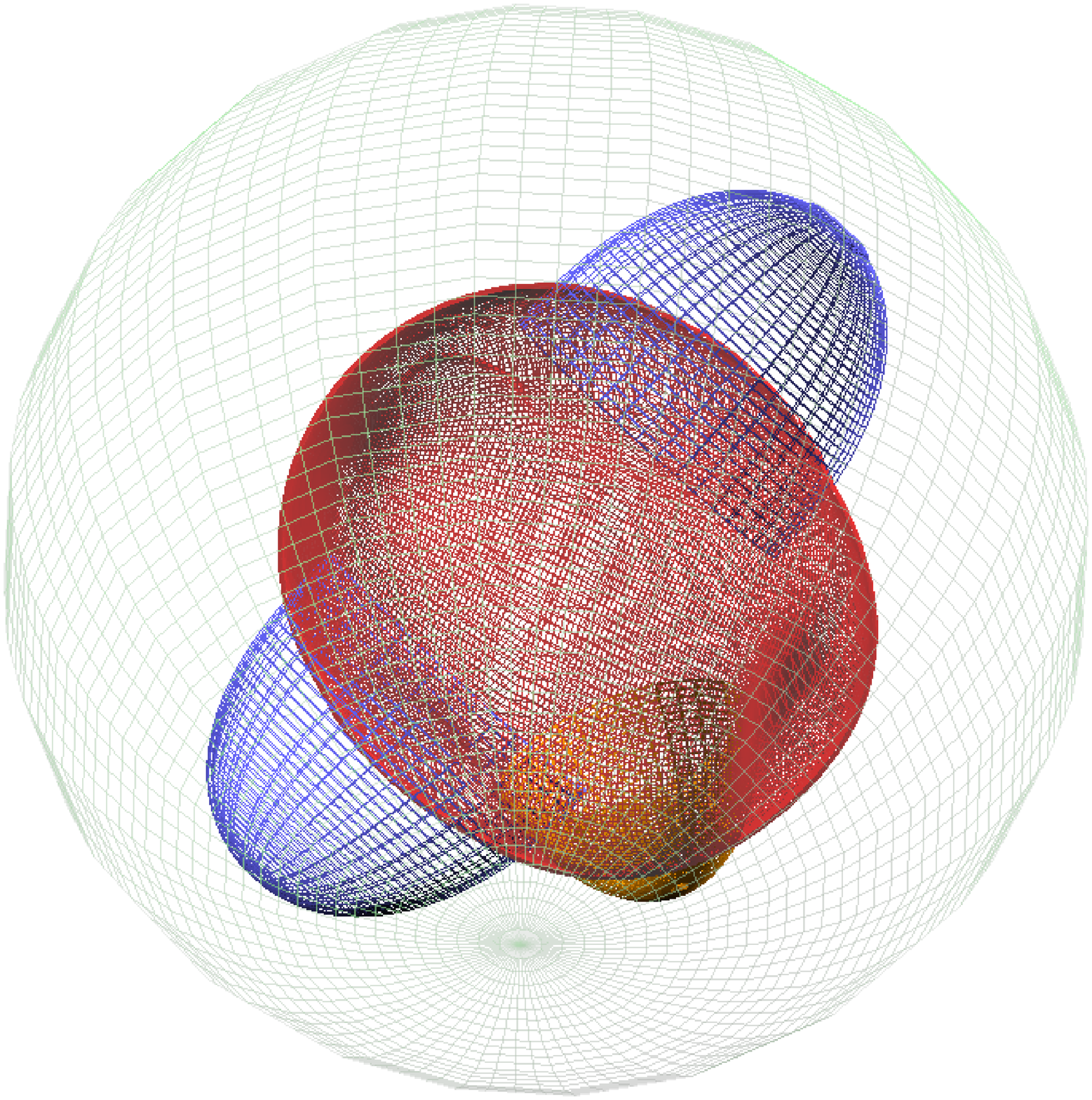}
\includegraphics[width=0.32\textwidth]{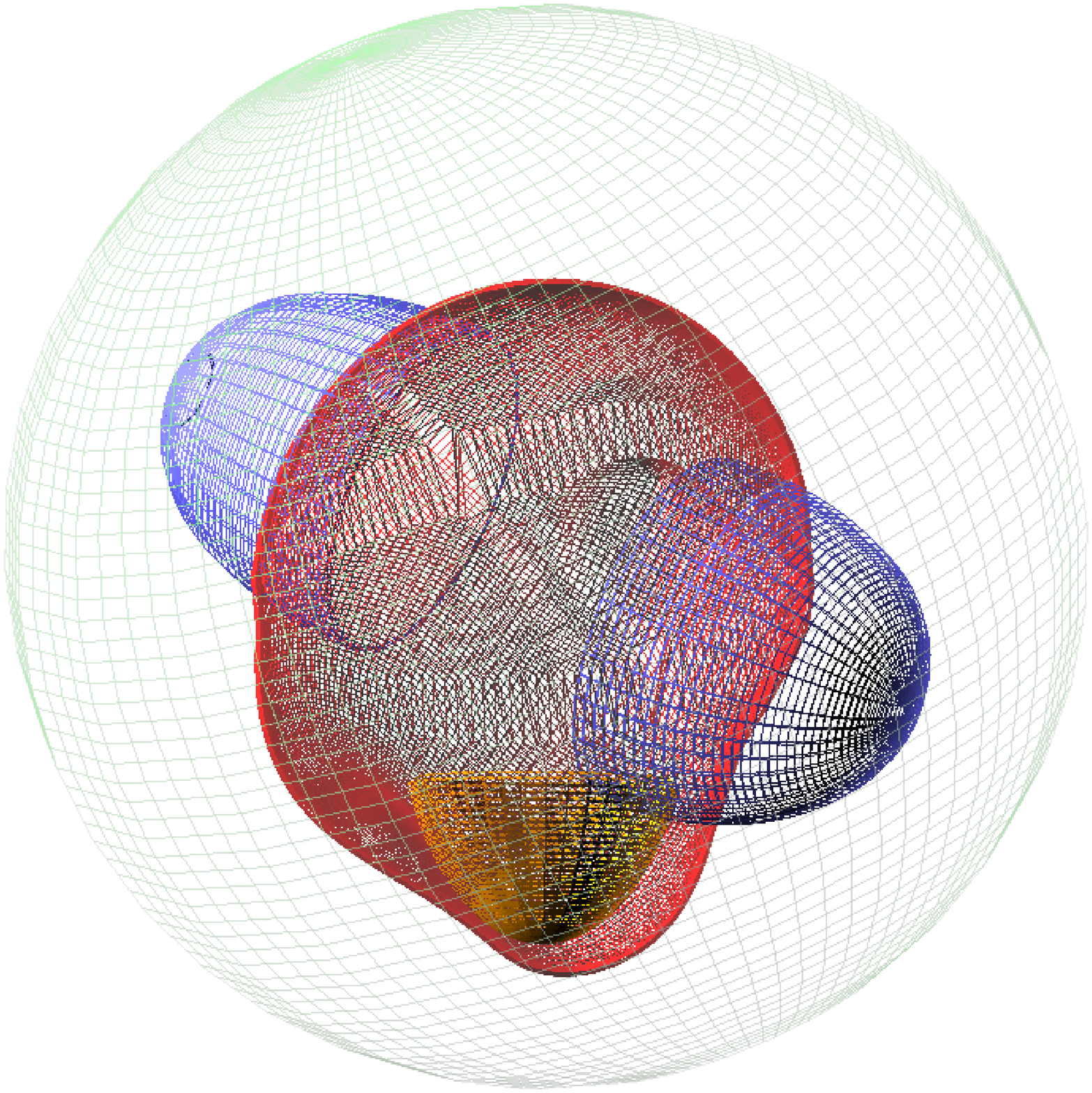}
\includegraphics[width=0.32\textwidth]{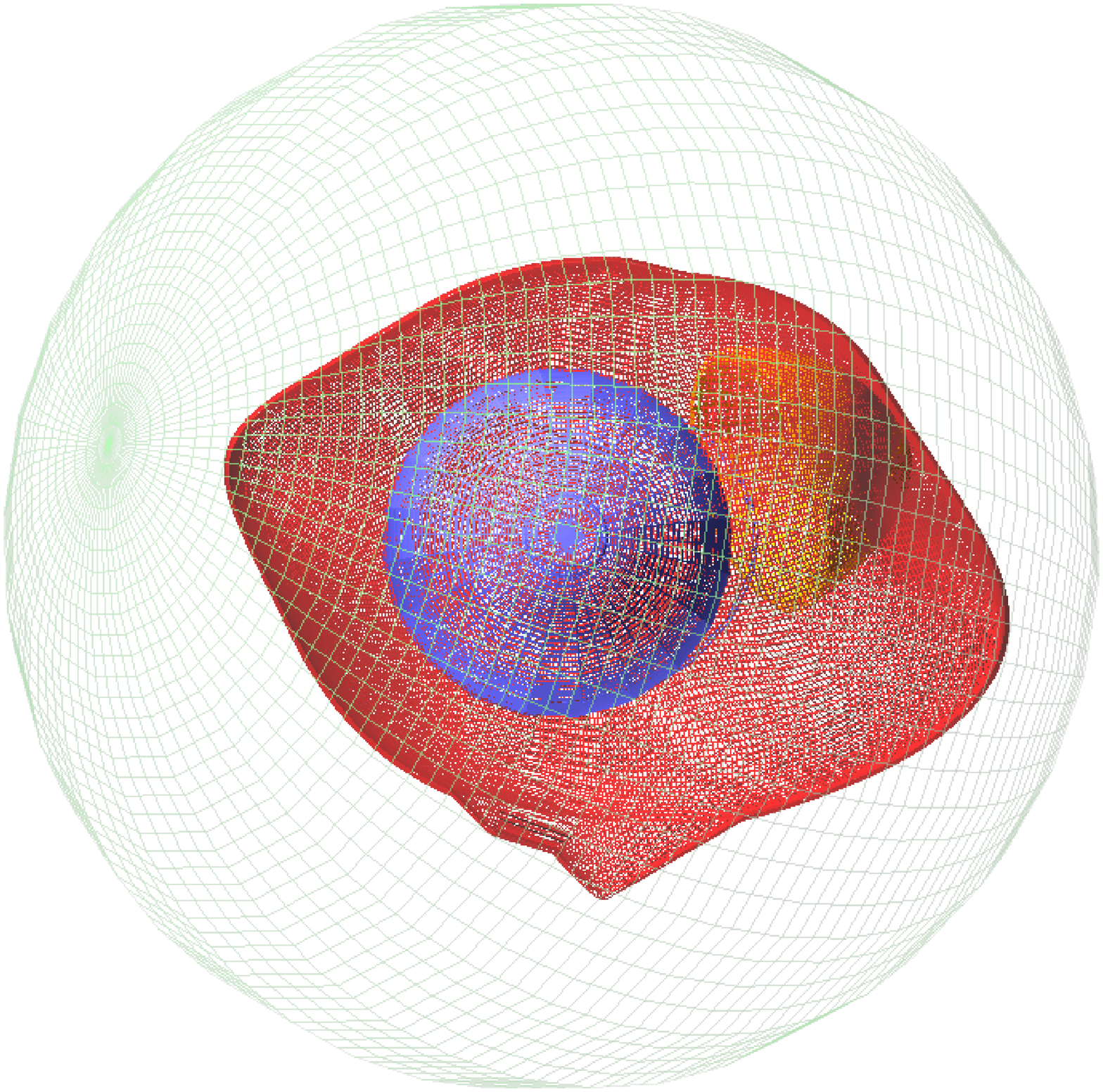}
\includegraphics[width=0.32\textwidth]{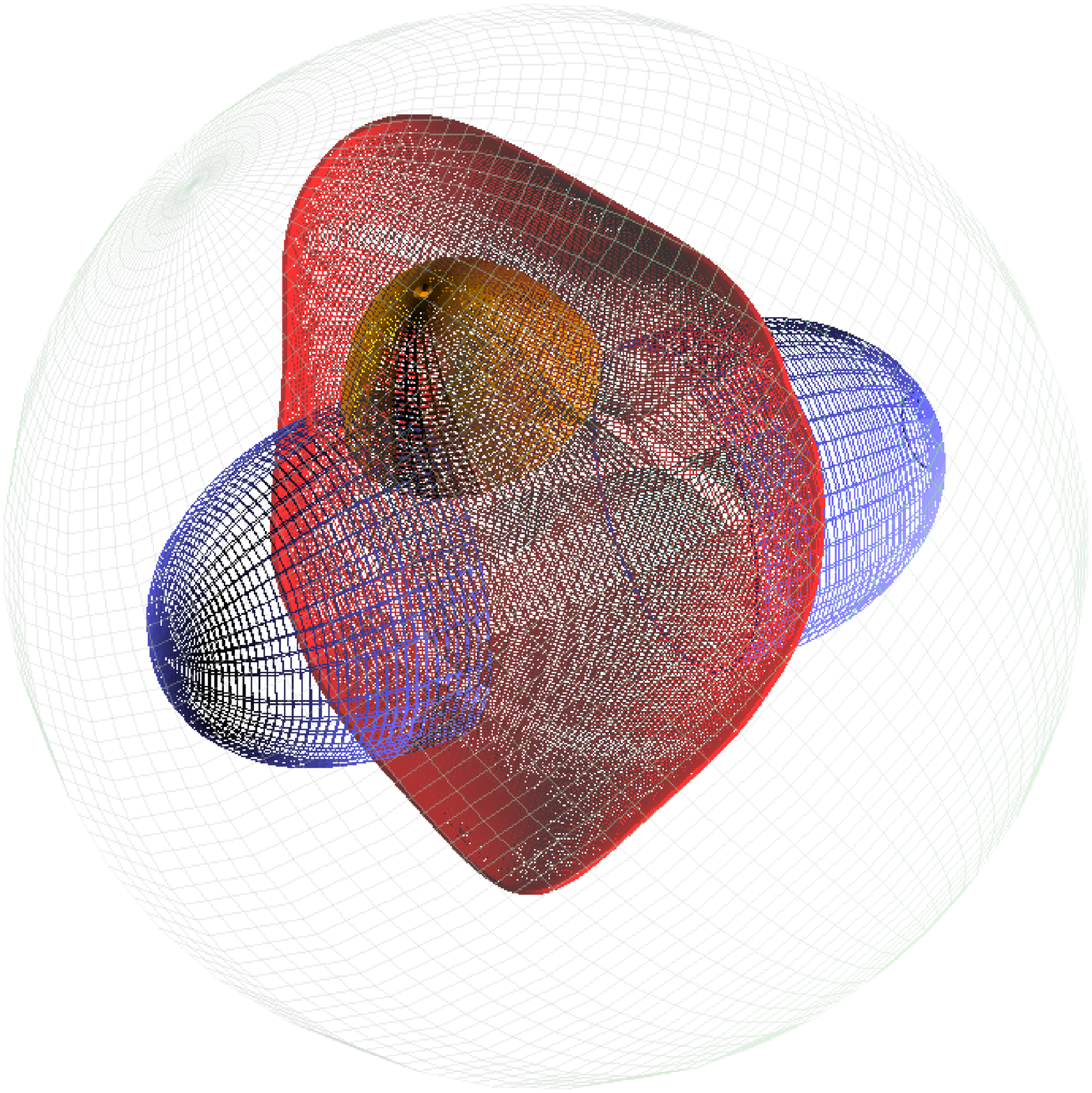}
\includegraphics[width=0.32\textwidth]{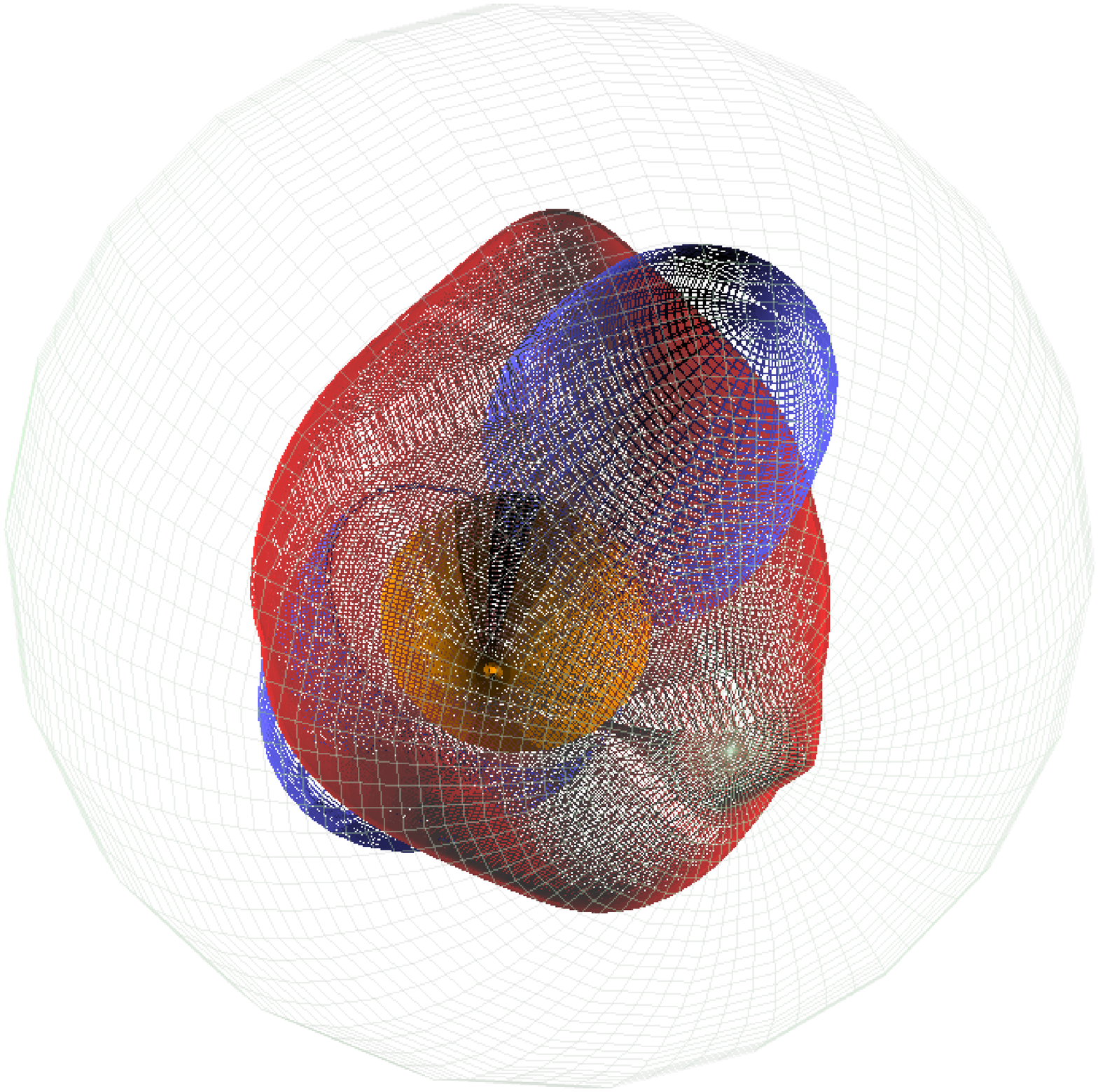}
  \vspace*{1pt}
  \caption{{\sc shape} model of NGC\,1514 from different perspectives. The top-left panel is the nebula as seen in 
  the plane of the sky (North up, East left). The rest of the views have been selected to clearly distinguish some 
  of the individual structures, namely. Top-center panel is tilted to show the full extension of bubbles B1a and B1b, 
  whereas bottom-left shows B1a in a pole-on view. Similarly, top-right panel is tilted to show the B2 bubble in 
  full extent and bottom-right shows B2 in the pole-on view. Finally, the bottom-center panel shows the full extent of the larger bump of the inner shell.}
  \label{fig:shape_3d}
\end{figure*}

\section{Kinematical reconstruction}
\label{sect:shape}

\subsection{Ionized nebula}

To try to understand the multiple components observed in the PV maps, we used the software {\sc shape} \citep{Steffen2011}, a three-dimensional modeling tool, in order to reproduce the main morphological structures and substructures in NGC\,1514. This software allows us to propose structures, assign them a velocity law, and vary their parameters to adjust both the image and the spectra (PV maps), interactively.  

We start with two spheres, one of them represents the faint outer shell of the nebula, whereas the other one
corresponds to the inner shell. This last structure was distorted by some modifiers called ``bumps'' that make
a protuberance that can be controlled with some parameters related to size and the direction of the protrusion. Three other structures were modeled as the tips of ellipsoidal shells, given the morphology seen in the 
images, but such structures are used as auxiliary tools to fit images and spectra. 

A sketch based on the final {\sc shape} model is shown in Figure~\ref{fig:sketch},
and main parameters of the structures are listed in Table~\ref{parameters}. Figure~\ref{fig:shape_spectra} shows a comparison between the observed PV-maps of NGC\,1514 and the synthetic PV-maps generated in {\sc shape} and in Fig~\ref{fig:shape_3d} the different structures of the {\sc shape} model are seen at different viewing angles. The main structures that fit image and PV diagrams are described, namely:

Outer Shell (OS). Although there is a tentative ellipticity from the PV map at S1, the evidence is not significant and the ellipticity would be very low. Therefore, this does not make a substantial difference with respect to the spherical model we propose for the outer shell.
This structure expands radially at 39\,km\,s$^{-1}$, following a velocity law of $V [{\rm km\,s^{-1}}] = 0.4 \theta$, being $\theta$ the angular radius in arcsec.

Inner Shell (IS): A distorted (although originally spherical) shell forms the main structure of the nebula. 
Expansion velocity of the original spherical shell was found as 26\,km\,s$^{-1}$, which is in good agreement with previous determinations \citep{Sabaddin-Hamzaoglu1982, Muthu-Anandarao2003}. We needed up to seven ``bump'' modifiers to model 
protrusions and cavities and thus distorting the original shell. The assumed velocity law 
for this shell is $V = 0.55\theta$, the same as for the bubbles (structures B1a,b and B2; see below). This velocity law allows 
us to reproduce very well both high-resolution spectra and images.

Bubbles 1 (B1a, B1b): Two bubbles are mostly symmetrically located in the nebula
and they were modeled as the polar regions of a prolate ellipsoid. The NW (SE) bubble
is labeled as B1a (B1b), and is located at PA$=-53\degr$ (127\degr) with an
inclination angle of $i=27\degr$ ($i$; with respect to the plane of the sky), being therefore blueshifted (redshifted).
The deprojected polar velocity of these bubbles is $V_{\rm pol}=42$\,km\,s$^{-1}$.

Bubble 2: Another bubble was modeled as the SW region, aligned at the same PA as 
one of the bumps ($-159\degr$), but with different kinematics. Deprojected polar velocity of this structure
was $V_{\rm pol}=39$\,km\,s$^{-1}$. 

The equatorial expansion velocity and radius of the different structures (see Table~\ref{parameters}), along with the distance of 454$\pm$4\,pc from the $Gaia$ $Early$ $Data$ $Release$ $3$ \citep[GEDR3;][]{GaiaCollaboration2020}, yield a kinematical ages (assuming homologous expansion, i.e., velocity proportional to radius) of, approximately, 5410\,yr (OS), 4000\,yr (B1a, B1b), 3750\,yr (B2) and 3970\,yr (IS).  All these estimates have uncertainties of $\sim300$ yr. This would mean that IS, B1 and B2 are, indeed, coeval estructures, as already expected having all of them the same velocity law. However, this result has to be taken with caution since, although it is true that the spherical shell model with bubbles seems to be appropriate, it is not possible to fully constrain the velocity law for each individual structure (or even to be sure that there is a unique velocity law for each structure), existing therefore a possible degeneracy. This is a usual problem in morpho-kinematical analysis. For this reason, in spite of the similar kinematical ages derived above, we propose in Section\,\ref{formation} different scenarios that may explain the formation of the bubbles and the other structures.

\subsection{Rings model}

As expected, there is no evidence of the mid-infrared rings of NGC\,1514 in our high-resolution optical spectra. If present, they have to be extremely faint. This is consistent with the conclusion by \cite{Ressler2010} that the nebula seems to contain a significant quantity of ionized gas while the rings do not. As already noted by \cite{Ressler2010}, the rings of NGC\,1514 closely resemble similar structures observed in other nebulae like the PN MyCn\,18 \citep{Clyne2014, Sahai1999} or the inner regions of the symbiotic nebula Hen\,2-104 \citep[also known as the Southern Crab;][]{Santander-Garcia2008}. Both have confirmed binary systems in their nuclei which are most likely responsible for the ring-like structures around them. Similar ring-like structures are seen in Abell\,14 \citep{Akras2016} although, in this case, the binarity of the central source has been proposed but not confirmed yet \citep{Akras2020}. Finally, the rings of NGC\,1514 are also comparable to the structures present in SN1987A \citep{Martin1995, Sugerman2005}. All of the mentioned examples are visible at optical wavelengths, unlike the rings detected in NGC\,1514, which are only seen in the infrared. In most of these cases, the basic structure to describe the rings is a sort of hourglass with a tight waist and open rims.  

\begin{figure*}
\includegraphics[width=0.92\textwidth]{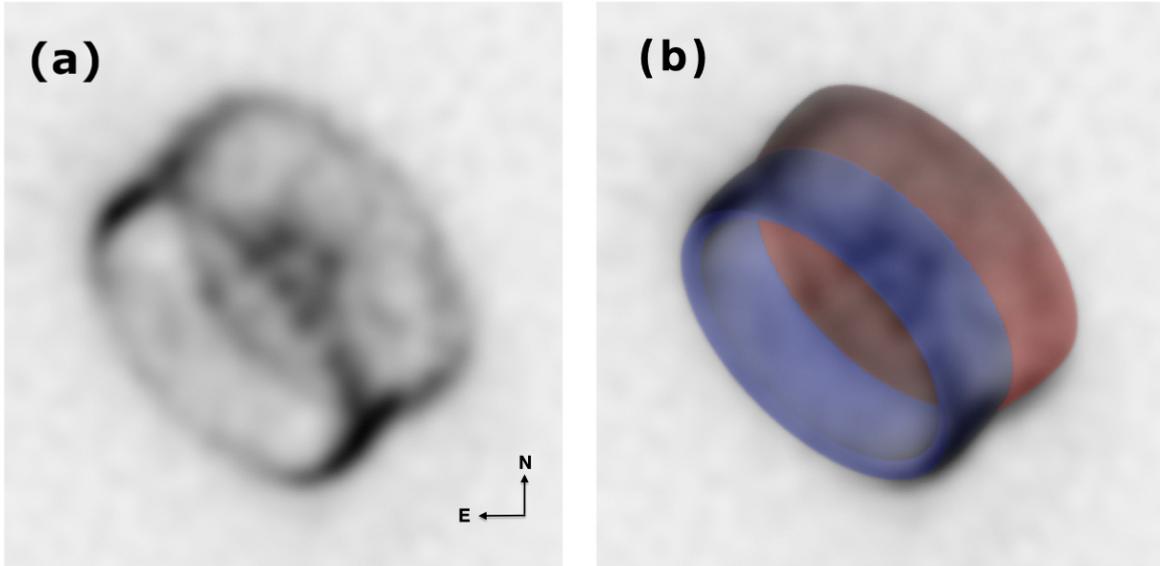}
  \vspace*{1pt}
  \caption{(a) Deconvolution result from the WISE W4 image (W4$_{\rm decon}$). (b) Barrel-like {\sc shape} model superimposed over the W4$_{\rm decon}$ image, with the SE outflow (ring) to the blue and the NW outflow (ring) to the red.}
  \label{fig:barrel}
\end{figure*}

With this in mind, we attempted to reproduce the rings seen in NGC\,1514 by using {\sc shape}. This is certainly challenging since we do not have information about the kinematics of the rings, a key point to constrain their morphology and inclination. Fig.\,\ref{fig:barrel}a shows the result of applying a deconvolution process \citep{Masci-Fowler2009} to the WISE W4 image (named as W4$_{\rm decon}$), where the rings are clearly recognized. As it can be seen in this image, the two ellipses formed by the rings are slightly different (with equal semi-major axis but different semi-minor axis). Therefore, if we assume that such ellipses are the two borders of an open bipolar outflow (we can think, for example, in a diabolo-like structure), the difference in size could be interpreted as a different inclination angle for one outflow respect to the other. This is certainly plausible but we could more easily reproduce the rings with a simpler geometry using a barrel-shaped structure with a pinched waist (or, similarly, a diabolo with a broader waist) and with its axis at $i$=28$^{\circ}$, as illustrated in Fig.\,\ref{fig:barrel}b. In this geometry, we do not need to invoke different orientations for the outflows, since the east part of the NW outflow would correspond indeed to the waist of the barrel-like structure. In addition, the NE and SW brightest regions of the rings are quite well explained with this structure instead of a pure diabolo model. Given the lack of kinematic information of the barrel-like structure, its orientation can not be easily interpreted, being the one showed in Fig.\,\ref{fig:barrel}b (with the SE outflow to the blue and the NW outflow to the red) just one possibility.

\section{Possible formation scenarios}
\label{formation}

The different structures seen in the the PV maps of NGC\,1514 would indicate that several ejection processes
may have been involved in the formation of the nebula. However, this would apparently clash with the kinematical ages derived for the inner shell and the bubbles (see Sect\,\ref{sect:shape}). Nevertheless, and as mentioned before, to constrain the velocity law of each structure is definitively a really difficult task, leading to a possible degeneracy. Therefore, we only propose here several scenarios that could have happened during the mass ejection. 

We can presume that the outer shell was the first structure to be formed, and may represent the spherical AGB wind from the star progenitor. Then, the dusty rings might have formed although, without their kinematical information, it is really difficult to put constraints on their formation (indeed, they could have even preceded the formation of the ionized nebula). A bipolar outflow would have sculpted the bubbles B1a and B1b. Finally, a subsequent spherical ejection could have formed the distorted inner shell. The visible deformation of this shell could be consequence of the interaction of the fast isotropic wind with the material of inhomogeneous density already present in the nebula. In this scenario, the wind is able to go through the low density regions of the nebula but is blocked by the high density regions, deforming the initial structure and forming the bubbles we see in the image \citep[see][]{Steffen2013}. Another possibility is that the inner (originally spherical) shell would have been formed after the outer shell and the dusty rings. Later, the ejection of a bipolar outflow (B1a and B1b) would have deformed the inner shell.
 	
As already mentioned, the nucleus of NGC\,1514 is a long-period ($\sim$ 9 years) binary system with an eccentricity of $\sim$ 0.5 \citep{Jones2017}. The influence of such wide binaries in PN morphologies is still uncertain (mainly due to the few PNe known to host long-period binaries, a consequence of the observational difficulties). In fact, hydrodynamical simulations show that the mass-transfer from an Asymptotic Giant Branch (AGB) star to its wide companion could also play a role in the shaping process of the nebula \citep[see e.g.,][]{Theuns1996, Edgar2008}. In the case of NGC\,1514, the highly aspherical shape and, more importantly, the presence of the pair infrared rings could be evidence of the impact of the binary central star on the formation of the nebula.

\begin{figure*}
\includegraphics[width=1.0\textwidth]{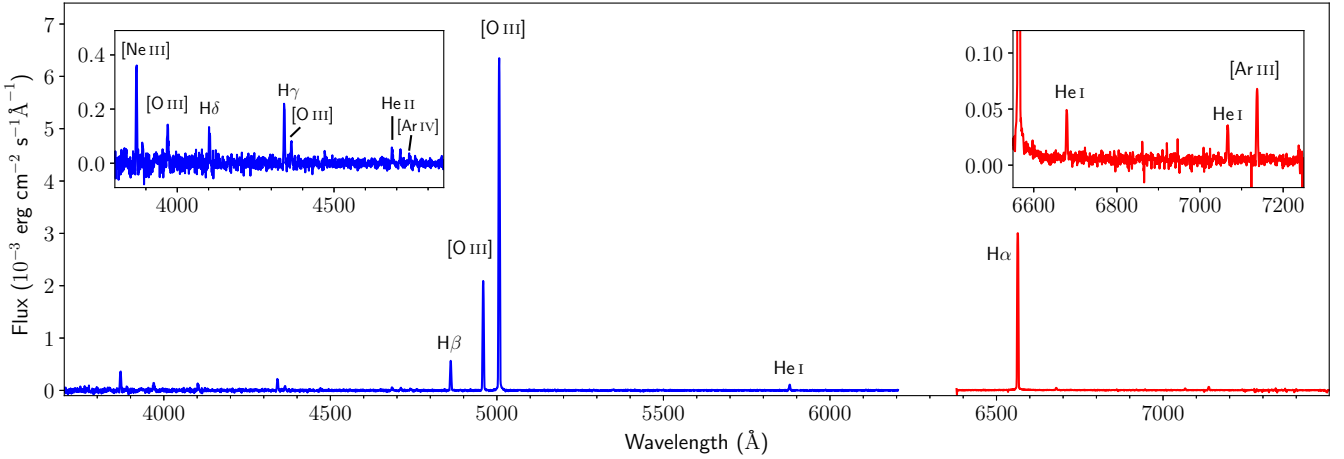}
  \vspace*{1pt}
  \caption{Goodman/SOAR spectrum of NGC\,1514 in the spectral range 3700 -- 7500 $\AA$. The gap between 6200 -- 6400 $\AA$ approximately, corresponds to the lack of data between the Blue and the Red modes of the grating. The two insets show the spectrum in the ranges 3800 -- 4850 $\AA$ and 6550 -- 7250 $\AA$ to highlight some of the faintest emission lines. For a complete identification of the emission lines see Table\,1).}
  \label{goodman-spectra}
\end{figure*}

In this context, we have investigated the scenario in which the complex morphology (and particularly the rings) may be the result of asymmetries in the mass-loss process of the primary star when the binary system was on a (hypothetical) shorter orbital period. If in this shorter orbit both components did not interact, then as a result of the mass-loss of the primary, the orbit could have increased. In order to numerically evaluate this scenario, we have calculated the initial separation of the binary system from the current separation ($a$ $\sim$ 20 au, being $a$ the semi-major axis) and assuming the masses of both stars published in \cite{Jones2017}. With these numbers, we derived an initial separation of $a$ $\sim$ 3 au \citep{Villaver2007}. We can in principle exclude the CE interaction on the basis of the current period of the system ($\sim$ 3300 days), since the final period of a CE interaction would have been at least an order of magnitude shorter than the current period of the system and it is hard to conceive an CE interaction the result of which is a system with the eccentricity observed in the binary pair of NGC\,1514. However, even without undergoing a CE evolution, with an initial separation of 3 au the system could still suffer significant interaction and mass transfer via, for example, wind Roche lobe overflow \citep{Theuns1996}, which allows high accretion rates. Indeed, some interaction mechanism like this is needed, since single models do not succeed in keeping the necessary surface rotational velocity for magnetic mechanisms to operate and create bipolar morphologies \citep{Garcia-Segura2016}.

Due to the relatively high eccentricity of the system, another possibility for the formation of the rings is that they were triggered by a periastron passage of the companion. This may be a plausible scenario if we take into account the numbers reported in \cite{Boffin-Jones2019} for this to occur. Unfortunately, the lack of kinematic information about the rings, prevents us to properly evaluate this scenario as done, for example, in \cite{Kashi2010} for Eta Carinae.
 
Finally, it is worthy to mention that the complex morphology of the nebula also lead \cite{Bear-Soker2017} to propose NGC\,1514 as one of the PNe shaped by a triple central star, although we do not have any evidence that supports such idea.

\section{Physical conditions and chemical abundances}
\label{sect:phys_cond}

For the analysis of the physical parameters and chemical abundances we used the Goodman spectrum. The position of the Goodman slit, at 30\,arcsec south from the central star, is shown in Figure\,\ref{fig:image_slits} (pink line). We have analyzed a region of 193.7\,arcsec containing the whole nebula (including the outer shell). In addition, we have also extracted separately the outer shell region (by adding both west and east parts) to check for the contribution of this outer shell to the inner one, and found that only H$\alpha$, H$\beta$ and [O\,{\sc iii}]$\lambda$$\lambda$4959,5007 are presented, as well as a very faint He\,{\sc ii} $\lambda$5876 emission line in the west part of it. These lines only represent the 5\% of the total flux measured on the whole nebula (see below).
Other small regions along the inner shell were also extracted, checking that the line intensity ratios and physical conditions were similar to those found for the whole region. Therefore, we will only describe the results obtained from the whole region in the following. 

Figure\,\ref{goodman-spectra} shows the intermediate-resolution, long-slit spectra of NGC\,1514 in the whole region. Apart from the strong H$\alpha$, H$\beta$, and [O\,{\sc iii}]$\lambda$$\lambda$4959,5007 emission lines, we can identify 
other faint nebular emissions, like [Ar\,{\sc iii}], [Ne\,{\sc iii}], and
He\,{\sc i} emission lines. Also, we identify a very faint He\,{\sc ii} $\lambda$4686 emission line, suggesting moderate excitation. Although the lack of a deeper and high-resolution, long-slit spectrum prevents us from making a complete and detailed chemical analysis of the nebula, we can extract some conclusions from the present spectra.

 We used the nebular software {\sc Anneb} \citep{Olguin2011} to compute the physical and chemical conditions of NGC\,1514. {\sc Anneb} integrates the {\sc nebular} package of {\sc iraf/stsdas} \citep{Shaw-Dufour1995}. The {\sc nebular} version used is the one contained in {\sc iraf} 2.16. From the fluxes of the identified emission lines in the spectrum, {\sc Anneb} calculates the logarithmic extinction coefficient $c$(H$\beta$) and 
both the electron temperature ($T_{\rm e}$) and density ($N_{\rm e}$) in an iterative way. It starts by assuming a case B recombination ($T_{\rm e}$=10$^{4}$\,K, $N_{\rm e}$=10$^{3}$\,cm$^{-3}$) and a theoretical H$\alpha$/H$\beta$ ratio of 2.85 \citep{Osterbrock-Ferland2006} as a first step. After convergence, it derives the dereddened line intensities and recalculates the final values of $c$(H$\beta$), $T_{\rm e}$, and $N_{\rm e}$ again (from different intensity ratios). Several extinction laws can be used in {\sc Anneb} and a proper error propagation is performed. 
  
For the calculations of $N_{\rm e}$, the most used emission lines ratios are [S\,{\sc
  ii}]$\lambda$6716/$\lambda$6731 and [O\,{\sc ii}]$\lambda$3729/$\lambda$3726. However, the lack of these emission lines in the spectrum of NGC\,1514 prevents us from deriving $N_{\rm e}$ from those ratios. For that reason, we have used the [Ar\,{\sc iv}]$\lambda$4711/$\lambda$4740 ratio, also used for the calculation of $N_{\rm e}$. But we have to take with caution this ratio since the [Ar\,{\sc iv}]$\lambda$4711 emission line is blended with the He\,{\sc i} $\lambda$4713 line in our spectrum, so we have estimated the contribution of the He\,{\sc i} $\lambda$4713 line in the observed fluxed and corrected this contribution according to the ratio $I$(4713)/$I$(4471)=0.154 \citep{Benjamin1999}.

 \begin{table}
\caption{Emission line intensities in NGC\,1514. The logarithmic extinction coefficient $c$(H$\beta$) and the physical parameters (electron temperature and density) are also indicated below.}            
\label{table:1}      
\centering           
\begin{tabular}{lrc}   
\hline  
 Line & $f(\lambda)$ &  \multicolumn{1}{c}{I$_{\lambda}$ }   \\   
\hline

[Ne\,{\sc iii}] $\lambda$3869          & 0.223    	&   102 $\pm$ 16 	\\

He\,{\sc i} $+$ H8 $\lambda$3889	    & 0.219     	&	19.0 $\pm$ 6.8		\\

H$\epsilon$ $+$ [Ne\,{\sc iii}] $\lambda$3970          & 0.202    	&	48 $\pm$ 11	\\

H$\delta$ $\lambda$4101              & 0.171    	&    31.2 $\pm$ 7.6   		\\	

H$\gamma$ $\lambda$4340          & 0.129    	&  51.4 $\pm$ 6.9 		\\	

[O\,{\sc iii}] $\lambda$4363    		& 0.124    &   19.6 $\pm$ 4.2		\\	

He\,{\sc i} $\lambda$4471          &  0.096    	&    5.5 $\pm$ 2.3 		\\	

He\,{\sc ii} $\lambda$4686          &  0.042    	&    12.5 $\pm$ 2.7 		\\	

He\,{\sc i} $+$ [Ar\,{\sc iv}]  $\lambda$4711          &  0.036    	&    8.2 $\pm$ 2.0 		\\

[Ar\,{\sc iv}] $\lambda$4740    & $0.029$     	&	 6.6 $\pm$ 2.8   	\\

H$\beta$ $\lambda$4861          & 0.000    		&  100.0 $\pm$ 6.8  		\\

[O\,{\sc iii}] $\lambda$4959    & $-0.024$    	&   355 $\pm$ 19 	\\

[O\,{\sc iii}] $\lambda$5007    & $-0.035$     	&    1064 $\pm$ 54 	\\

He\,{\sc i} $\lambda$5876          & $-0.217$    	&   12.7 $\pm$ 1.5 		\\

H$\alpha$ $\lambda$6563         & $-0.323$    	&   280 $\pm$ 20 	\\

He\,{\sc i} $\lambda$6678          & $-0.339$     	&  3.59 $\pm$ 0.63  		\\

He\,{\sc i} $\lambda$7065          & $-0.383$     	&  2.43 $\pm$ 0.50  		\\

[Ar\,{\sc iii}] $\lambda$7136    & $-0.391$     	&	 4.97 $\pm$ 0.72  		\\

\hline

log$F$$_{\rm H\beta}$(erg\,cm$^{-2}$\,s$^{-1}$) &         &	$-12.597$ $\pm$ 0.021 \\

$c$(H$\beta$)  			&	&   0.876 $\pm$ 0.067     \\

$T_{\rm e}$ ([O\,{\sc iii}]) ($K$) &	&   14800 $\pm$ 2200   \\

$N_{\rm e}$ ([Ar\,{\sc iv}])$^{a}$ (cm$^{-3}$)	&	&   3200 $\pm$ 1100     \\

\hline    

\label{fig:emission_lines}                              
\end{tabular}

$^{a}$ For the calculation of $N_{\rm e}$ from the [Ar\,{\sc iv}]$\lambda$4711/$\lambda$4740 intensity ratio, we have estimated and subtracted the contribution of He\,{\sc i} $\lambda$4713 to the observed [Ar\,{\sc iv}]  $\lambda$4711 emission line (see the text).

\end{table}

Table\,\ref{fig:emission_lines} lists the emission line intensities and
their Poissonian errors. The emission line fluxes have been dereddened with the corresponding derived $c$(H$\beta$) (see bottom part of the table) and the extinction 
law $f$($\lambda$) of \cite{Seaton1979}. {$N_{\rm e}$ and $T_{\rm e}$ are also included in the table. Neither $N_{\rm e}$ nor $T_{\rm e}$ shows substantial fluctuations along the nebula. $T_{\rm e}$ is in agreement with that derived from other authors previously \citep{Manchado1989, Milanova-Kholtygin2009}. In contrast, our derived electron density is much higher than that previously reported by \cite{Kohoutek1967}, who obtained $N_{\rm e}$ = 290 cm$^{-3}$ for the inner shell based on the H$\beta$ flux and assuming a spherical shell with a uniform distribution of radiating matter. Likewise, \cite{Ressler2010} inferred an electron density below 10$^{3}$ cm$^{-3}$ based on the [S\,{\sc iii}] 18.7/33.5 $\mu$m line ratio, also considerably lower than our result.

From the outer shell spectrum we derived a mean $c$(H$\beta$)$\sim$0.85, similar to the whole nebula (see Table\,\ref{fig:emission_lines}) although it is slightly larger ($\sim$0.98) in the west part than in the east one ($\sim$0.72). Fig.\,\ref{fig:profiles} shows the surface brightness profiles of the [O\,{\sc iii}]$\lambda$5007 and H$\alpha$ emission lines along the slit. It is noticeable that the outer shell brightness profile follows a roughly linear dependence with the angular distance, as expected from a type II multiple-shell PN with an attached shell \citep{Chu1987,Guerrero1998}. Due to the few emission lines detected in the outer shell, no physical conditions or chemical abundances can be derived for this region.

\begin{figure}
\includegraphics[width=0.45\textwidth]{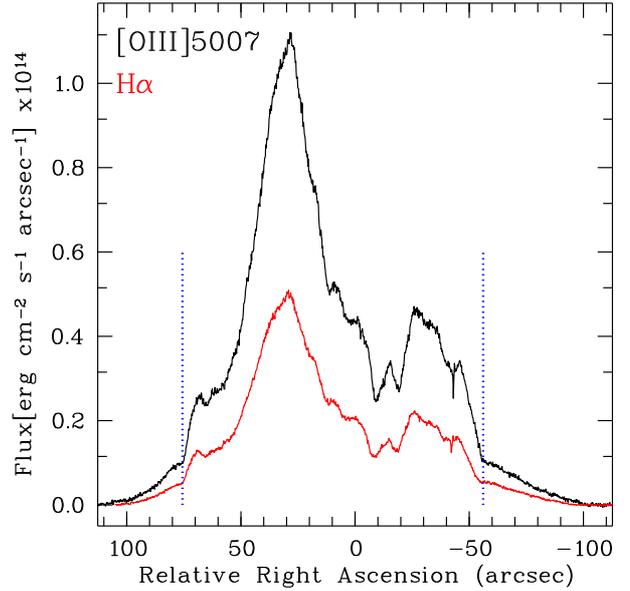}
  \vspace*{1pt}
  \caption{Intensity profiles of the [O\,{\sc iii}]$\lambda$5007 (black) and  H$\alpha$ (red) emission lines along the slit. The origin of the angular distance corresponds to the right ascension of the central star and the blue dotted lines correspond to the frontiers between the inner shell and the outer shell.}
  \label{fig:profiles}
\end{figure}

Ionic abundances of the whole nebula covered by our slit are listed in Table\,\ref{tab:abundances}.
For those ions with more than one line observed in the spectrum, 
the reported ionic abundance value was calculated as a weighted average 
using as weight the signal-to-noise ratio of the line. Obtaining the elemental abundances is not immediate, mainly due to the fact that not all the stages of ionization are seen for all the elements. Therefore, we only can provide a lower limit for most of the elements (argon, neon, and oxygen), since we only have one stage of ionization (two in the case of the argon, where abundance has been calculated as a sum of the individual ionic abundances of Ar$^{2+}$  and Ar$^{3+}$). Finally, the helium abundance was calculated by using the method of \cite{Kwitter-Henry2001}. The results are listed in Table\,\ref{tab:abundances}. The values obtained are in agreement with previous abundance determinations from other authors \citep{Manchado1989, Costa2004, Stanghellini2006}. In contrast to other previous authors \citep[see, e.g.,][]{Milanova-Kholtygin2009,Manchado1989}, we are not able to derive a nitrogen abundance.

Finally, for comparison purposes, we have also calculated the chemical abundances of NGC\,1514 using the code {\sc PyNeb} \citep[v. 1.1.14;][]{Luridiana2015}. The obtained ionic and elemental abundances are listed in Table\,\,\ref{tab:abundances} and, as can be compared, all of them are in very good agreement with those obtained with {\sc Anneb}. Regarding the physical conditions, similar values to those derived with {\sc Anneb} are also obtained in {\sc PyNeb}, being $T_{\rm e}$ ([O\,{\sc iii}]) = 14500$\pm$1500\,K and $N_{\rm e}$ ([Ar\,{\sc iv}]) =  2800$\pm$1800\,cm$^{-3}$.

 \begin{table}
\caption{Ionic (top) and elemental (bottom) abundances of NGC\,1514 derived with \sc{Anneb} and \sc{PyNeb}. }            
\label{table:1}      
\centering           
\begin{tabular}{lcc}   
\hline  
Ratio                                             &     Abundance  (\sc{Anneb})         &     Abundance  (\sc{PyNeb})    \\
\hline
He$^{+}$/H$^{+}$ &   (8.41$\pm$0.92)$\times$$10^{-2}$  &		(7.8$\pm$2.1)$\times$$10^{-2}$\\
He$^{2+}$/H$^{+}$   &   (1.01$\pm$0.22)$\times$$10^{-2}$ &	(1.08$\pm$0.25)$\times$$10^{-2}$	\\
Ne$^{2+}$/H$^{+}$   &   (2.68$\pm$0.41)$\times$$10^{-5}$ &	(2.9$\pm$1.5)$\times$$10^{-5}$\\
O$^{2+}$/H$^{+}$    &   (1.162$\pm$0.043)$\times$$10^{-4}$ &	(1.21$\pm$0.30)$\times$$10^{-4}$\\
Ar$^{2+}$/H$^{+}$   &   (2.06$\pm$0.30)$\times$$10^{-7}$ &		 (1.88$\pm$0.65)$\times$$10^{-7}$ \\
Ar$^{3+}$/H$^{+}$   &   (5.2$\pm$1.1)$\times$$10^{-7}$ &		(6.5$\pm$4.2)$\times$$10^{-7}$\\
\hline 
He/H	 	        	 & 0.094$\pm$0.010     	&	0.089 $\pm$0.021\\
Ne/H$^{a}$  	& $>$ (2.68$\pm$0.41)$\times$$10^{-5}$	& 	$>$ (2.9$\pm$1.5)$\times$$10^{-5}$\\
O/H	$^{a}$     & $>$ (1.162$\pm$0.043)$\times$$10^{-4}$ 	& $>$ (1.21$\pm$0.30)$\times$$10^{-4}$\\
Ar/H$^{b}$ 	& $>$  (7.9 $\pm$1.2)$\times$$10^{-7}$  		&	$>$  (8.38$\pm$4.3)$\times$$10^{-7}$ \\
\hline    
\end{tabular}
\label{tab:abundances}  

$^{a}$ These are lower limits since they have been calculated with an only ionization state.
$^{b}$ Argon abundance calculated as a sum of the individual ionic abundances of Ar$^{2+}$  and Ar$^{3+}$.

\end{table}

We would like to note that a deeper and higher resolution spectrum is required for doing a comprehensive and detailed chemical analysis both from the optical recombination lines and collisionally excited lines.

\section{Conclusions}

We present, for the first time, a detailed morpho-kinematical analysis of the complex planetary nebula NGC\,1514 based on optical high-resolution, long-slit spectra. From the position-velocity (PV) maps and with the help of the software {\sc shape}, we generated a 3D model of NGC\,1514, which provides information on the main structural components as well as their kinematics. We propose an inner distorted shell (originally spherical) with some additional and well-defined bubbles (two of them mostly symmetrically located in the nebula) and an outer spherical attached shell. Except for this diffuse outer shell that was the first component to be ejected, the rest of the structures might follow the same velocity law and, therefore, the kinematical ages of those structures would be apparently the same.

The two large mid-infrared rings detected by \cite{Ressler2010} in the WISE images are not identified in our high-resolution spectra, which prevented us from doing a kinematical analysis of them. Nevertheless, in view of their resemblance to similar structures in other nebulae as, e.g., the PNe MyCn\,18, Abell\,14, or the symbiotic nebula Hen\,2-104, we propose a barrel-like structure with an inclination of $i$=28$^{\circ}$ (based only in the morphological information) to explain the rings.

Finally, we also present an analysis of the physical parameters and chemical abundances of the nebula by means of intermediate-resolution, long-slit spectra. The nebular spectra reveal a moderate-excitation nebula with emission lines of [Ar\,{\sc iii}], [Ne\,{\sc iii}], He\,{\sc i} and He\,{\sc ii} (and also weak [Ar\,{\sc iv}]). 
An electron temperature of $\sim$15000\,K and an electron density between 2000 and 4000\,cm$^{-3}$ have been obtained for the nebula, the latter being substantially higher than previously reported values.

\section*{Acknowledgements}

We are very grateful to our referee, Martin A. Guerrero, for his valuable comments which have improved the presentation of the paper. Also, we would like to thank Eva Villaver and David Jones for the very useful discussion on the evolutionary status of the binary system. AA acknowledges support from Government of Comunidad Aut\'onoma de Madrid (Spain) through postdoctoral grant `Atracci\'on de Talento Investigador' 2018-T2/TIC-11697 and support from FONDECYT through postdoctoral grant 3160364. R.V. acknowledges support from UNAM-PAPIIT IN106720,
LFM acknowledges partial support by MCIU grant AYA2017-84390-C2-1-R,
co-funded with FEDER funds, and financial support from the State Agency
for Research of the Spanish MCIU through the ``Center of Excellence
Severo Ochoa" award for the Instituto de Astrof\'isica de Andaluc\'ia
(SEV-2017-0709). The work of MER was carried out at the Jet Propulsion Laboratory, California Institute of Technology, under a contract with NASA.
 This research has made use of the SIMBAD database, operated at the CDS, Strasbourg (France), Aladin, NASA's
Astrophysics Data System Bibliographic Services. This research has made use of the Spanish Virtual Observatory (http://svo.cab.inta-csic.es) supported from the Spanish MINECO/FEDER through grant AyA2017-84089. This research has been partly funded by the Spanish State Research Agency (AEI) Project MDM-2017-0737 at Centro de Astrobiolog\'ia (CSIC-INTA), Unidad de Excelencia Mar\'ia de Maeztu. Authors also acknowledge the Calar Alto Observatory for the service observations. Finally, authors also acknowledge support from OAN-SPM staff, in particular to Mr. Francisco Guill\'en for his assistance during night operations.

\section*{Data availability}

The data underlying this article will be shared on reasonable request to the corresponding author.




\bibliographystyle{mnras}
\bibliography{Bibliography} 





%


\bsp	
\label{lastpage}
\end{document}